\documentclass[12pt]{iopart}

\usepackage{graphicx}
\usepackage{iopams}

\begin{document}

\title{Microcanonical treatment of black hole decay at the Large Hadron
Collider}

\author{Douglas M Gingrich\footnote{Also at TRIUMF, Vancouver, BC V6T
2A3 Canada.} and Kevin Martell}

\address{Centre for Particle Physics, Department of Physics, University
of Alberta, Edmonton, AB T6G 2G7 Canada}

\ead{\mailto{gingrich@ualberta.ca}, \mailto{martell@ualberta.ca}}

\begin{abstract}
This study of corrections to the canonical picture of black hole decay
in large extra dimensions examines the effects of back-reaction
corrected and microcanonical emission at the LHC.
We provide statistical interpretations of the different multiparticle
number densities in terms of black hole decay to standard model
particles.
Provided new heavy particles of mass near the fundamental Planck scale
are not discovered, differences between these corrections and thermal
decay will be insignificant at the LHC. 
\end{abstract}


\section{Introduction\label{sec1}}

In higher-dimensional, low-scale gravity
scenarios~\cite{Arkani98,Antoniadis,Arkani99,Randall99a,Randall99b},
black holes can be produced in high-energy particle collisions such as
those anticipated from the Large Hadron Collider
(LHC)~\cite{Banks,Giddings01a,Dimopoulos01}.  
Most of these models predict that black holes will rapidly decay in four
successive phases of which Hawking evaporation~\cite{Hawking74,Hawking75}
may dominate.
Accordingly, the Hawking evaporation phase has been the most studied
decay phase in higher-dimensional models.   

The thermodynamic description of black
holes~\cite{Bekenstein73,Bardeen,Bekenstein74,Hawking76} has been a
powerful tool for probing quantum gravity aspects of black hole physics.   
The statistical mechanical interpretation of Hawking evaporation treats
the black hole as a constant temperature reservoir that allows the
emission of particles in thermal equilibrium with the black hole. 
While this is probably a good approximation for large black holes, its 
applicability to small primordial black holes or higher-dimensional
black holes in low-scale gravity is not clear.

Modelling emissions with the canonical ensemble from these types of
black holes is inappropriate since a black hole in asymptotically flat
spacetime cannot be in stable thermal equilibrium with its radiation. 
Consequently, when the mass of the black hole is close to its
temperature, its evaporation will be modified. 
The Hawking evaporation picture can be improved by applying the laws of 
statistical mechanics to ensure that the evaporation remains
thermodynamically valid below the Planck scale.
This can be done by using back-reaction corrected emission in a
microcanonical ensemble 
\cite{Harms92,Harms93a,Harms95a,Harms95b,Hossenfelder04p,Casadio01a}.  
In addition, Ref.~\cite{Alberghi06p} has provided an alternative
implementation of the back reaction of the emitted radiation through a
modification of the relation between the black hole radius and its
temperature. 

The effects of quantum gravity should become important at the mass
scales we will consider.
Since the thermodynamic description of quantum systems is often a useful
tool, the microcanonical description of black hole evaporation can be
expected to remain valid in the quantum regime near the Planck scale.  

This paper is organized as follows.
Sec.~\ref{sec2} reviews Hawking evaporation.
Following this, Sec.~\ref{sec3} introduces the microcanonical ensemble
and back-reaction corrected emissions.
It also discusses the regime in which the microcanonical description
becomes important.
Sec.~\ref{sec4} addresses the measurable particle energy spectra.
Sec.~\ref{sec5} argues that, at the LHC, the microcanonical corrections
will probably not be important. 
The paper concludes with a discussion of the applicability of the
modified energy spectrum if new heavy particles are emitted from black
holes.   

\section{Hawking Evaporation\label{sec2}}

Hawking radiation is not described by a pure quantum-mechanical state
but by a density matrix. 
It is completely thermal in that the probabilities of emission of
particles in different modes and probability of emitting different
number of particles in the same mode are completely uncorrelated.
The probability for different numbers of particles agree exactly with
thermal radiation.
It is assumed in Hawking evaporation that the black hole emits
non-interacting particles.

The expectation value of the number of particles $\langle N \rangle$ of
a given species emitted in a mode with frequency $\omega$, angular
momentum $m$ about the axis of rotation of the black hole, and charge
$q$ is  

\begin{equation} \label{eq1}
\langle N \rangle =
\frac{\Gamma(\omega)}{\rme^{2\pi(\omega-m\Omega-q\Phi)/\kappa} + s}\, ,  
\end{equation}

\noindent
where $s$ is a statistics factor that is $-1$ for bosons and $+1$ for
fermions;  
$\kappa$, $\Omega$, and $\Phi$ are the surface gravity, surface angular
frequency of rotation, and surface electrostatic potential;
the species dependent $\Gamma(\omega)$ is the fraction of the mode that
would be absorbed were it incident on the black hole.
Since the expected number of particles emitted in each mode is the same
as that of a thermal body whose absorptivity matches that of the black
hole, the temperature of the black hole has been identified as $T = 
\kappa/(2\pi)$.  

Hawking evaporation can be described by a number density.
The number density represents the available states that can be occupied
by Hawking radiation. 
For a non-rotating and non-charged black hole, the number density is
most often represented by the canonical ensemble number density

\begin{equation} \label{eq2}
n_T(\omega) = \frac{1}{\rme^{\omega/T} + s}\, ,
\end{equation}

\noindent
where $\omega$ is the energy of the emitted particle (usually assumed to
be massless) and the frequency-dependent (grey-body) factor $\Gamma$ has
been ignored for now.
There are also charged and rotating black hole solutions, which
contribute to the grand canonical ensemble in which the electronic
potential and rate of rotation act as the chemical potential for
charge and angular momentum~\cite{Gibbons}.

The canonical ensemble is a hypothetical collection of systems of
particles used to describe an actual individual system, which is in
thermal contact with a heat reservoir but not allowed to exchange
particles with its environment. 
The black hole is treated as a heat bath of fixed temperature and the
back reaction of the particles on the black hole metric is neglected.  

Because the temperature of a black hole increases as the mass decreases,
it cannot be in stable thermal equilibrium.  
In addition, in asymptotically flat spacetime, black holes have a
negative specific heat thereby implying that the canonical ensemble does    
not apply. 
Hawking emission of particle energies close to the black hole mass have
non-negligible back reactions.  
This back reaction is responsible for modifying the temperature.
Thus the heat-bath assumption breaks down as the mass of the black hole 
approaches the Planck scale.
The back reaction of the metric has not yet been completely solved.

For black holes with mass of the order of the Planck scale, or less, the 
thermodynamical picture is no longer a good approximation of the true
microcanonical description. 
We can make the above description valid from a thermodynamic point of
view, even when the energy of the emitted particle is close to the black
hole mass, by using the microcanonical ensemble.

The statistical mechanical description of black holes does not provide
any time information.
Two extreme views can be taken.
At one extreme, all the emissions can be considered to occur so quickly
that the black hole mass, and thus temperature, is approximately
constant at its initial value during the decay. 
This is the so called ``sudden approximation'' view take by Dimopoulos
and Landsberg~\cite{Dimopoulos01}. 
The other extreme view is that the decay takes a long time and the black
hole reaches thermal equilibrium after each emission before emitting the
next particle.
This quasistationary picture is adapted throughout this paper. 

\section{Microcanonical Ensemble\label{sec3}}

To describe black hole decay, we will employ the microcanonical ensemble
of a large number of similar insulated systems each with a given fixed
energy. 
Each of these systems will then have a number of different
configurations compatible with the given energy. 
These configurations then form a shell in the configuration space of the
system. 
As time passes, the system moves from one point to another in this
shell.
Then by the assumption of ergodicity, the probability of the whole
system being a particular one in a chosen region of configuration space
is proportional to the number of configurations in that region whose
energies lie within an infinitesimal range.

The microcanonical ensemble number density for a single-particle
microstate is given by the exponential of the change in black hole
entropy before and after the emission 

\begin{equation} \label{eq3}
n_\mathrm{SP}(\omega) = \rme^{-\Delta S} =
\frac{\rme^{S(M-\omega)}}{\rme^{S(M)}}\, , 
\end{equation}

\noindent
where $S(M)$ is the entropy of the black hole with initial mass $M$ and
$S(M-\omega)$ the entropy of the black hole after emitting a particle of
energy $\omega$.
This single-particle distribution can be understood by interpreting the 
occupation of states as arising from a tunnelling
probability~\cite{Parikh}. 
The distribution has also been derived as the emission rate from excited
$D$-branes using the microcanonical ensemble with back-reaction
corrected emission rates in field theory~\cite{Keski-vakkuri,Das}.  

From the single-particle number density, the average particle density
can be obtained by counting the multiplicity of states according to
their statistics~\cite{Koch05}:

\begin{equation} \label{eq4}
n_\mathrm{BR}(\omega) = \frac{1}{\rme^{S(M) -S(M-\omega)} +s}\, ,
\end{equation}

\noindent
where $\omega < M$.
This distribution approaches the thermal distribution (\ref{eq2})
for $\omega \ll M$. 
The number density has similarities to an ideal gas of thermal radiation
in equilibrium with a fix-temperature heat bath.
The $\omega/T$ argument in the exponential of the canonical
multiparticle number density has essentially been replaced by $\Delta
S$. 
$\Delta S$ has the nice property of always being bounded between 0 and
1.
$\Delta S$ is the change in entropy of the black hole due to a single
particle being in a state with mode energy $\omega$.
The $k$-th particle state is defined by a change in entropy, not in
energy value.  
A $k$-particle state changes the entropy by $k\Delta S$, but does not
have an energy $k\omega$.
For fixed $M$ and $\omega$, the change in entropy is the constant
quantity that sets the spacing between adjacent modes.
More than one particle in a state must be viewed as moving to that state
simultaneously before the black hole mass can change. 
For a black hole, the mass and entropy cannot change in such a way as to
keep the temperature fixed.
In other words, the mass and temperature of the black hole change in
such a way as to keep $\Delta S$ constant. 
The energy of the black hole plus particle system is conserved.
Thus the distribution represents the back-reaction corrected number
density of an ideal gas.
We refer to (\ref{eq4}) as the multiparticle distribution for
back-reaction corrected emissions in an ensemble of particle-occupied
excited modes.
The ideal gas analogy should not be pushed too far as the equivalent $q$
or $\psi$ function to the canonical or grand canonical ideal gas do not
allow a straight forward determination of the macroscopic properties of
the gas. 

Equation~(\ref{eq4}) is not the only possibility for the multiparticle
number density. 
By considering a black hole as an extended quantum object
($p$-brane), which is made of other black holes, and then considering a 
gas of $p$-brane black holes~\cite{Casadio98,Cirilo}, the occupation
number density for the Hawking particles in the microcanonical ensemble
has been proposed as 

\begin{equation} \label{eq5}
n_E(\omega) = \sum_{k=1}^\infty \frac{\Omega(E-k\omega)}{\Omega(E)}
\Theta(E-k\omega)\, .
\end{equation}

\noindent
In thermodynamic equilibrium, the statistical mechanical density of
states is $\Omega(M) = \rme^{S(M)}$.  
By identifying the total energy of the system with the initial black
hole mass, we write

\begin{equation} \label{eq6}
n_E(\omega) = \sum_{k=1}^{\lfloor \frac{M}{\omega} \rfloor}
\rme^{S(M-k\omega)-S(M)}\, , 
\end{equation}

\noindent
where $\lfloor M/\omega \rfloor$ denotes the integer part of $M/\omega$.
This distribution approaches (\ref{eq4}) and thus the thermal
distribution (\ref{eq2}) for $\omega \ll M$. 
Now each $k$-particle state is defined by a mode energy and changes the
black hole entropy by a different amount.
The number of particles in a state is not determined by the particle
statistics but rather by truncating the sum appropriately to conserve
energy.
Each term can be viewed as a state but the number of particles in each
state is not fixed.
For example, one particle in a state with energy $k\omega$ is the same
as $k$ particles in a state with energy $\omega$.
Since each term can be viewed as a different energy of emission, and
thus entropy change, $\omega$ represents the quanta of energy or mode 
energy. 
When plotting $n_E(\omega)$ versus $\omega$ we are allowing the
mode energy $\omega$ to vary. 
The plot built up by scanning through the mode energies is the same as
scanning through the particle energies. 
We refer to (\ref{eq6}) as the multiparticle distribution of
back-reaction corrected emissions in a microcanonical ensemble of
$p$-brane defined black holes.
Since (\ref{eq6}) does not include the emitted particle's statistics
factor $s$, it must be added by hand or the equation can only be applied
for large changes in entropy, where the evaporation is governed by
Boltzmann statistics.    

Equation~(\ref{eq6}) has been used to calculate the black hole
lifetime~\cite{Hossenfelder04p,Casadio01a,Casadio00}.  
In four dimensions, the evaporation rate $\rmd M/\rmd t$ diverges at
$M=0$ if the canonical number density (\ref{eq2}) is used.
The canonical ensemble justifies the use of the sudden approximation.
On the other hand, using the microcanonical number density
(\ref{eq6}) leads to a finite decay rate and gives a lifetime for
black holes with $M = 2M_\mathrm{P}$ that is $10^9$ times longer than
that given by the canonical number density~\cite{Hossenfelder04p}.
In higher dimensions, $\rmd M/\rmd t$ is finite and slows down in the
later stage of evaporation.
The microcanonical ensemble justifies the use of the quasistationary
approximation. 

We can understand the relationship between (\ref{eq4}) and
(\ref{eq6}) a bit better by restoring the units in (\ref{eq6}) (as
pointed out by Casadio and Harms~\cite{Casadio98}) to get

\begin{equation} \label{eq7}
n_E(\omega) = \sum_{k=1}^{\lfloor \frac{Mc^2}{\hbar\omega} \rfloor}
\rme^{S(Mc^2-k\hbar \omega)-S(Mc^2)}\, .
\end{equation}

\noindent
In the classical limit for fixed $c$ and $G_D$, $\hbar \rightarrow 0$
and the upper limit in the sum $Mc^2/\hbar\omega \rightarrow \infty$.
Only the lowest-order term proportional to $\omega/T$ remains when
expanding the argument of the exponential. 
Since $M_{P} \propto (\hbar c/G_D)^{1/(n+2)} \rightarrow 0$,
$\hbar \rightarrow 0$ is equivalent to $M/M_{P} \rightarrow \infty$.
Thus the finite sum becomes important at black holes masses near the
Planck scale, and (\ref{eq6}) can be considered the quantum version of 
(\ref{eq4}).

To compare the number densities, expand the entropy to leading order in
$\omega/M$ to obtain   

\begin{eqnarray} \label{eq8}
S(M-\omega) & = & S(M) - \frac{\omega}{T} - \frac{1}{2C_V} \left(
\frac{\omega}{T} \right)^2  + \cdots \nonumber \\ 
& \approx & S(M) - \frac{\omega}{T}\, ,
\end{eqnarray}

\noindent
where $C_V$ is the black hole specific heat.
Thus for $\omega \ll M$, the microcanonical number density approaches the
back-reaction corrected density and canonical density.

To examine the higher-order contributions, we need to write down the
entropy expressions.
For a black hole of mass $M$ and horizon radius $R$ in $n$
extra-dimensional asymptotically flat spacetime, the entropy is  

\begin{equation} \label{eq9}
S(M) = \frac{4\pi}{n+2} R M = \frac{n+1}{n+2} \frac{M}{T}
\end{equation}

\noindent
and the entropy differences is

\begin{equation} \label{eq10}
\Delta S = S(M) - S(M-\omega) = S(M) \left[ 1 - \left( 1 -
\frac{\omega}{M} \right)^\frac{n+2}{n+1} \right]\, . 
\end{equation}

\noindent
Assuming the maximum difference between the distributions occurs at the
highest energy emissions (in the kinematic limit $\omega \rightarrow M/2$) 

\begin{equation} \label{eq11}
(\Delta S)_\mathrm{max} \approx \frac{\omega}{T} - \frac{M}{2T} \left\{
1 - 2\frac{n+1}{n+2} \left[ 1 - \left( \frac{1}{2}
\right)^\frac{n+2}{n+1} \right] \right\}\, .  
\end{equation}

\noindent
The biggest difference from a thermal distribution occurs at low values
of $n$:

\begin{equation} \label{eq12}
(\Delta S)_\mathrm{max} \approx \frac{\omega}{T}- 0.05 \frac{M}{T}\quad 
\mathrm{for}\quad n = 2\, .
\end{equation}

\noindent
For large $M/T$ (large $S$),

\begin{equation} \label{eq13}
n_T(\omega) \rightarrow \rme^{0.05 M/T} n_\mathrm{BR}(\omega)\, .
\end{equation}

\noindent
The exponential in $n_\mathrm{BR}(\omega)$ dies quickly so the
correction factor in the numerator does not have much effect. 
For small $M/T$, the exponential is of the same order as $s=\pm 1$.
In this case, the correction factor will have a significant effect. 

\section{Particle Energy Spectrum\label{sec4}} 

Experimentally, we measure the energy spectra for different particle
types.
The energy spectra can be predicted from the number density as follows. 
The particle emission rate from a non-rotating and non-charged black
hole in three dimensions is

\begin{equation} \label{eq14}
\frac{\rmd N}{\rmd t} \propto n(\omega) \frac{\rmd^3k}{(2\pi)^3}\, .
\end{equation}

\noindent
For isotropic massless particle emission, 

\begin{equation} \label{eq15}
\frac{\rmd N}{\rmd \omega} \propto \omega^2 n(\omega)\, .
\end{equation}

\noindent
Often we consider the energy spectrum as a probability distribution and
normalize (\ref{eq15}) to unity over some region of energy.  
This is especially common for Monte Carlo generators like
CHARYBDIS~\cite{Harris03a,Gingrich06b}.  

In the rest frame of the black hole, conservation of energy-momentum
requires a particle with mass $m$ to be emitted with an energy $\omega$
in the range

\begin{equation}\label{eq16}
m < \omega \le \frac{M}{2} \left[ 1 + \left( \frac{m}{M}\right)^2
\right]\, . 
\end{equation}

\noindent
The canonical distribution (\ref{eq2}) does not respect
energy-momentum conservation and any value of $\omega$ is allowed.
Although the thermodynamic concept breaks down for $\omega \ge M$, the
distribution does not enforce this condition.
Throughout this paper, the minimum black hole mass is assumed to be
close to a Planck scale of about 1~TeV. 
The exact value of the Planck scale is not too important since the black
hole mass is expressed in terms of the Planck scale. 
However, the Planck scale can not be much higher than a few TeV if we
are to observe black holes at the LHC.  
Assuming the heaviest particle continues to be the top quark at LHC
energies, the largest value of the emitted particle energy will be only
3\% above the value $M/2$.
We thus neglect the particle mass and take the upper limit on the
emitted particle energy to be $M/2$.
This kinematic limit affects most of the decays and modifies the energy
spectrum of emitted particles in the canonical ensemble~\cite{Harris05a}. 

The definition of the Planck scale is important when considering
energies near the Planck scale. 
The PDG~\cite{PDG} definition of the Planck scale 

\begin{equation} \label{eq17}
M_D^{n+2} = \frac{(2\pi)^n}{8\pi G_D} 
\end{equation}

\noindent
has been chosen for use throughout this paper.
This definition causes the factor in the entropy that does not depend on
the ratio of the black hole mass to Planck scale to increase
monotonically with increasing number of dimensions.   
If the Dimopoulos and Landsberg definition is used, a minimum in the
entropy factor occurs at $n=3$~\cite{Gingrich06a}.
For the PDG definition of the Planck scale and seven extra dimensions,
the black hole mass approaches its temperature when it is about 0.3
times the Planck scale.  
In the following, cases for $n=2$ to 7 are studied, $n=2$ is referred to
as low dimension and $n=7$ as high dimension.

\section{Results\label{sec5}}

We now examine the affect the different number densities have on the
particle energy spectra. 
Table~\ref{boson} shows the maximum difference between the canonical and
back-reaction corrected energy spectra for different dimensions for
bosons. 
Table~\ref{fermion} shows the corresponding results for fermions.
The maximum differences range from about 10\% to 26\% and occur at a
black hole mass equal to the Planck scale for bosons and equal to about
1.3 times the Planck scale for fermions.
The maximum differences occur at a temperature of about 225~GeV and an
entropy of about 4. 
In all cases, except for four dimensional black holes emitting bosons,
the maximum difference occurs at temperatures well below the black hole
mass.  

\begin{table}[heb]
\caption{\label{boson}Maximum difference $\Delta$ between the canonical
and back-reaction corrected energy spectra for bosons.
$M$ and $T$ are in units of $M_D$ and $M_D = 1$~TeV.}
\begin{indented}
\item[]\begin{tabular}{@{}ccccccc}
\br
$n$ & $M$ & $T$ & $S(M)$ & $S(M/2)$ & $\Delta$ & $\Delta$ (\%) \\
\mr
2 & 0.987 & 0.264 & 2.80 & 1.11 & 0.0106 & 21 \\
3 & 0.973 & 0.240 & 3.25 & 1.37 & 0.0067 & 17 \\
4 & 0.981 & 0.230 & 3.55 & 1.55 & 0.0050 & 14 \\
5 & 1.001 & 0.227 & 3.78 & 1.68 & 0.0041 & 12 \\
6 & 1.023 & 0.227 & 3.94 & 1.78 & 0.0035 & 11 \\
7 & 1.049 & 0.229 & 4.07 & 1.86 & 0.0031 & 10 \\
\br
\end{tabular}
\end{indented}
\end{table}

\begin{table}[htb]
\caption{\label{fermion}Maximum difference $\Delta$ between the
canonical and back-reaction corrected energy spectra for fermions.   
$M$ and $T$ are in units of $M_D$ and $M_D = 1$~TeV.}
\begin{indented}
\item[]\begin{tabular}{@{}ccccccc}
\br
$n$ & $M$ & $T$ & $S(M)$ & $S(M/2)$ & $\Delta$ & $\Delta$ (\%) \\
\mr
2 & 1.394 & 0.235 & 4.44 & 1.76 & 0.0073 & 26 \\
3 & 1.336 & 0.221 & 4.83 & 2.03 & 0.0048 & 21 \\
4 & 1.323 & 0.217 & 5.08 & 2.21 & 0.0037 & 17 \\
5 & 1.333 & 0.217 & 5.28 & 2.36 & 0.0030 & 14 \\
6 & 1.352 & 0.218 & 5.42 & 2.45 & 0.0026 & 13 \\
7 & 1.379 & 0.222 & 5.53 & 2.54 & 0.0024 & 11 \\
\br
\end{tabular}
\end{indented}
\end{table}

\begin{figure}[htb]
\begin{center}
\includegraphics[width=12cm]{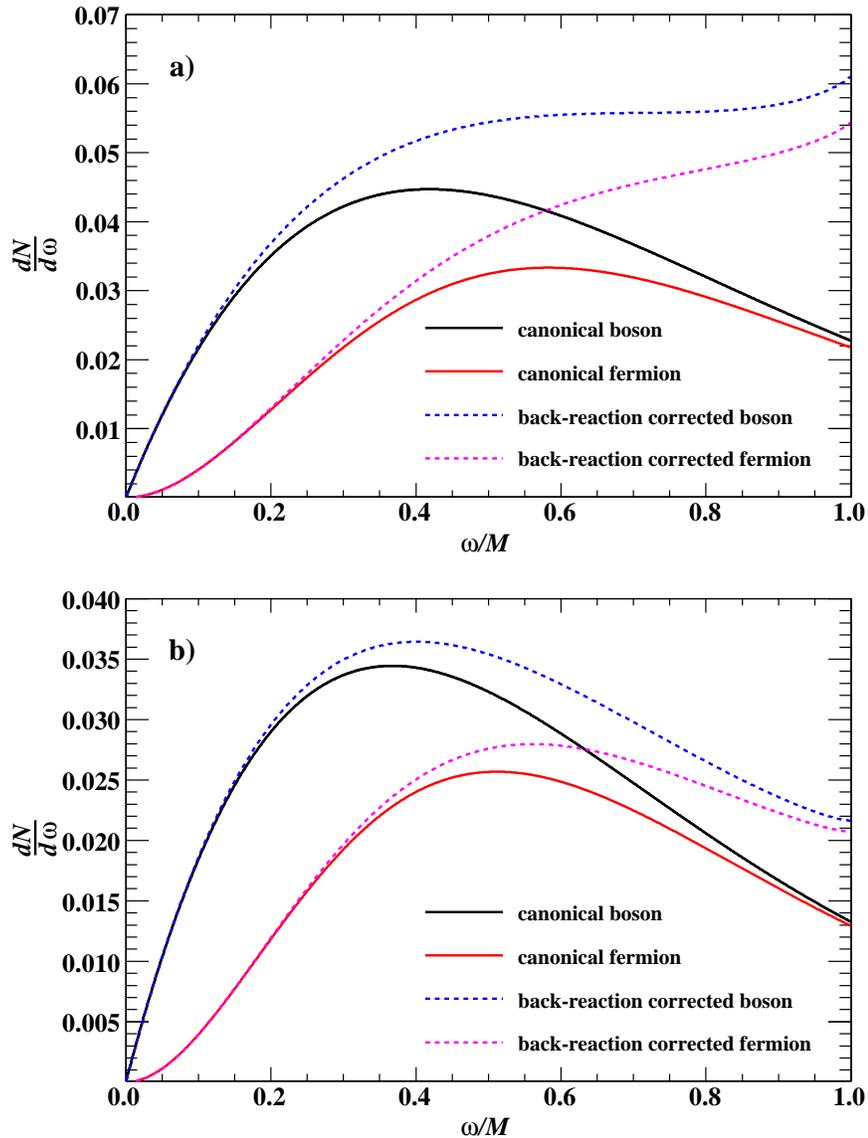}
\end{center}
\caption{Energy spectra for $M = M_D = 1$~TeV. 
a) $n=2$, $T=260$~GeV, $S=3$ and
b) $n=7$, $T=230$~GeV, $S=4$.
\label{fig1}}
\end{figure}

For large entropy, the canonical and back-reaction corrected energy
distributions are similar. 
For $M = 5M_D$, the distributions restrict the particle energies to be 
$\omega \lesssim 0.3M$, with
about $\omega \approx 0.06M$ being the most probable value.
As the black hole decays down to the Planck scale, we see notable
differences between the canonical and back-reaction corrected
distributions. 
Figure~\ref{fig1} shows the energy distributions as a function of
$\omega/M$ using the canonical number density (\ref{eq2}) and
back-reaction corrected number density (\ref{eq4}).
The normalization of the curves has not been modified from (\ref{eq2})
and (\ref{eq4}).
There is a significant difference between the distributions.
The distributions extend over the entire kinematic range and deviate
from each other at about $\omega \approx 0.2M$ for boson and $\omega
\approx 0.4M$ for fermions.
The biggest differences occur at $\omega = M$, with significant
differences still occurring at $\omega = 0.5M$.
The back-reaction corrected distributions (figure~\ref{fig1}a)) for low
dimensions favour energies at the highest kinematically allowed particle
energies.

\begin{figure}[htb]
\begin{center}
\includegraphics[width=12cm]{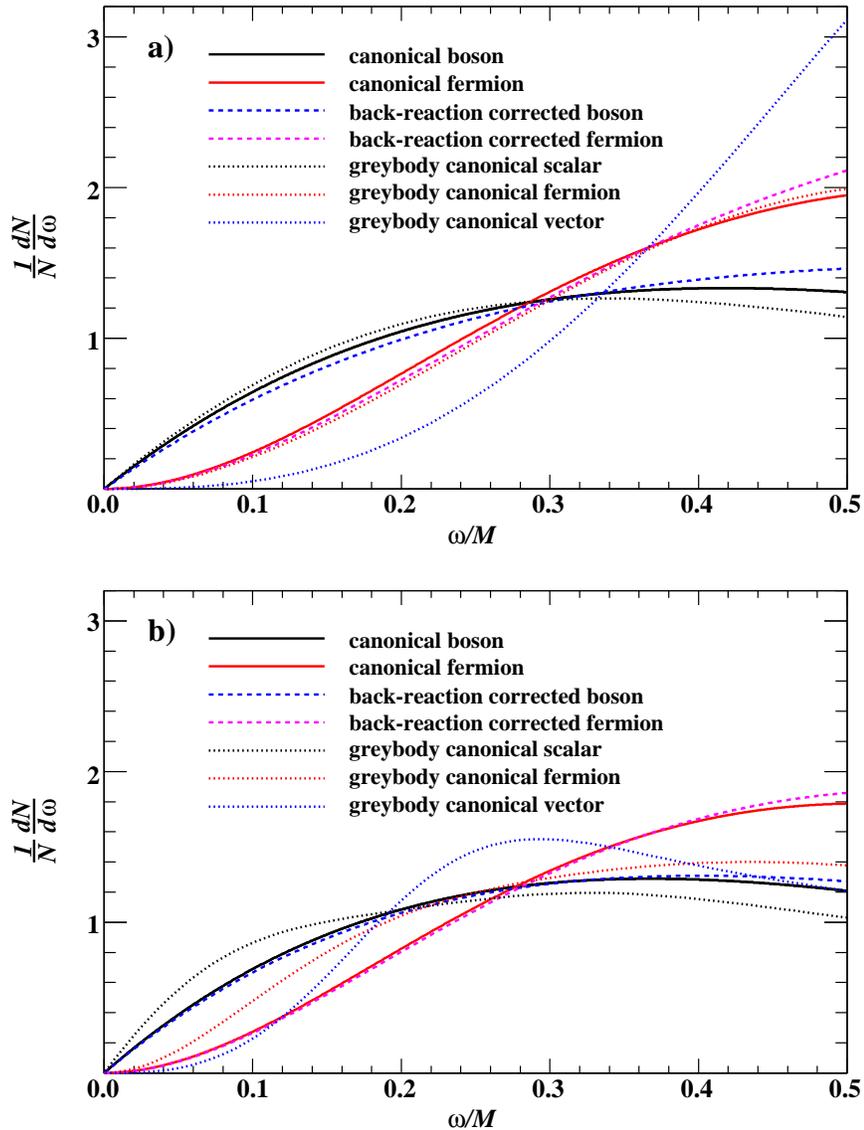}
\end{center}
\caption{Normalized energy spectra for $M = M_D = 1$~TeV. 
a) $n=2$, $T=260$~GeV, $S=3$ and
b) $n=7$, $T=230$~GeV, $S=4$.
\label{fig2}}
\end{figure}

Interpreting the energy distributions as probability density functions,
requires that the curves be normalized within the physically allowed
region of $0 < \omega \le M/2$, as shown in figure~\ref{fig2}.
There is now very little difference between the canonical and
back-reaction corrected distributions after normalization. 
For comparison, the greybody distributions using the canonical
number density have been included as the dotted
lines~\cite{Harris03a,Harris03b}. 
For higher dimensions, the effect of the greybody factors is more
significant than the back reaction correction.
In all cases, the vector boson greybody factors give a significant
difference.  
For the affects of greybody factors on high-entropy black holes see
\cite{gingrich07}.

Identical distributions to those presented throughout this paper have
been obtained using the black hole Monte Carlo event generator CHARYBDIS
by restricting the mass of the black hole to an arbitrary range of 2~GeV
around $M$ and looking at the energy of the first emitted particle in
the black hole rest frame. 
About $10^6$ events are needed to reduce the statistical fluctuations to
where the figures in this paper are effectively reproduced.
Small discontinuities in the distributions are visible at the mass
thresholds for the top quark and heavy gauge bosons.

\begin{figure}[htb]
\begin{center}
\includegraphics[width=12cm]{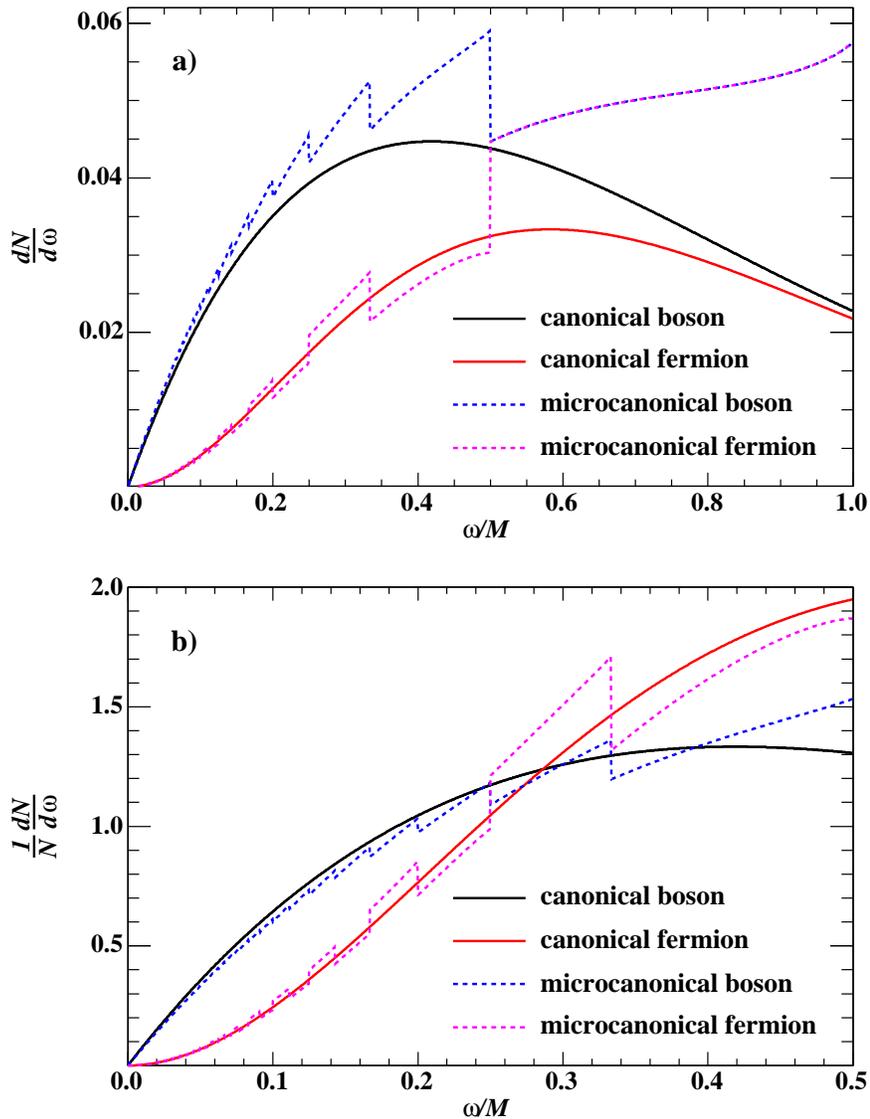}
\end{center}
\caption{a) Energy spectra and b) normalized energy spectra for 
$M = M_D = 1$~TeV, $n=2$, $T=260$~GeV, $S=3$.
\label{fig3}}
\end{figure}

Figure~\ref{fig3} shows energy distributions as a function of $\omega/M$ 
using the canonical number density (\ref{eq2}) and microcanonical
number density (\ref{eq6}). 
For black hole masses near the Planck scale, the finite sum's impact
becomes clear.
The microcanonical distributions are again more similar to the canonical
distributions after normalizing over the kinematic range of $\omega <
M/2$, as shown in figure~\ref{fig3}b).
In future experiments, finite detector acceptance and resolution effects
will most likely wash out the structure in the microcanonical
distributions.

\section{Discussion\label{sec6}}

The six previous distributions are rather theoretical.
At the LHC, the laboratory is unlikely to ever be in the rest frame of
the black hole. 
Not boosting the particle energy spectra back into the rest frame of the
black hole will wash out any signature of Hawking evaporation; boosting
the particles is essential. 
Since the effects here appear near the Planck scale, the entire decay
chain (time ordered) must be reconstructed or focus must be placed on
low-mass black hole production.
Studies of time ordering of black hole decay have been shown to not be
very effective~\cite{Harris03a,Palmer}. 
In addition, at the end of a decay chain the black hole becomes
highly boosted and thus the energy of particles in the laboratory frame
can be as high as $M$ to $2M$.
The other approach is to study only the first emitted particle from
low-mass black holes. 
In this case, the multiplicities are low and we are immediately in the
quantum gravity regime.
Equivalent studies of first-emitted particles for high-mass black holes
are far from trivial~\cite{Harris03a,Palmer}. 
Applying the techniques to black holes near the Planck scale should only
be attempted with a realistic simulation of a detector.

The differences between the canonical and microcanonical energy
distributions 
are very small over the allowed kinematic region for the Standard Model
particles emitted by black holes above the Planck scale.
Distinguishing between the ensembles will require an accurate
determination of the black hole four-momentum in order to boost the
particle energy into the black hole rest frame. 
Taking into account detector acceptance and resolution effects,
distinguishing between the canonical and microcanonical distributions 
is unlikely to be possible at the LHC.

If a new heavy particle is discovered to decay from black holes at the
LHC, it may be possible to identify the statistical ensemble provided
the energy spectrum of the particle can be measured.  
For example, for a heavy particle of about half the Planck scale, the
allowed kinematic region in black hole decay is $0.5 < \omega/M <
0.625$.  
In this region, the canonical and microcanonical energy distributions
are significantly different and normalization over a narrow region
of $\omega/M$ does not affect the difference.  
However, since decays to the heavy particle well above the Planck scale
do not uniquely identify the ensemble distribution, the decays must
be identified near the Planck scale, which could be problematic. 
In addition, the probability of emission of such a heavy particle would
be small so it would be hard to accumulate reasonable statistics for
such decays, even if they did occur. 

\ack

We would like to thank Don N. Page for helpful discussions.
This work was supported in part by the Natural Sciences and Engineering
Research Council of Canada.

\section*{References}

\end{document}